\documentclass[12pt]{iopart}
\usepackage{epsfig,graphicx,tabularx}
\usepackage{psfrag}
\usepackage{graphicx}
\usepackage{graphics}
\usepackage{epsfig}
\usepackage{bm}
\usepackage{verbatim,color,ulem}
\usepackage{amsfonts}
\usepackage{amssymb}

\newcommand{\beq}{\begin{equation}}
\newcommand{\eeq}{\end{equation}}
\newcommand{\beqa}{\begin{eqnarray}}
\newcommand{\eeqa}{\end{eqnarray}}

\newcommand{\be}{\begin{equation}}
\newcommand{\ee}{\end{equation}}
\newcommand{\bea}{\begin{eqnarray}}
\newcommand{\eea}{\end{eqnarray}}

\begin{document}

\title{Dephasing-rephasing dynamics of one-dimensional tunneling quasicondensates}

\author{A. Tononi}
\address{Dipartimento di Fisica e Astronomia 'Galileo Galilei', 
Universit\`a di Padova, via Marzolo 8, 35131 Padova, Italy}

\author{F. Toigo}
\address{Dipartimento di Fisica e Astronomia 'Galileo Galilei', 
Universit\`a di Padova, via Marzolo 8, 35131 Padova, Italy}

\author{S. Wimberger}
\address{Dipartimento di Scienze Matematiche, Fisiche ed Informatiche, 
Universit\`a di Parma, Parco Area delle Scienze 7/A, 43124 Parma, Italy}
\address{INFN, Sezione di Milano Bicocca, Gruppo Collegato di Parma, Parma, Italy}

\author{A. Cappellaro}
\address{Dipartimento di Fisica e Astronomia 'Galileo Galilei', 
Universit\`a di Padova, via Marzolo 8, 35131 Padova, Italy}

\author{L. Salasnich}
\address{Dipartimento di Fisica e Astronomia 'Galileo Galilei', 
Universit\`a di Padova, via Marzolo 8, 35131 Padova, Italy}

\date{\today}

\begin{abstract}
We study the quantum tunneling of two one-dimensional quasi-condensates
made of alkali-metal atoms, considering two different tunneling configurations:
side-by-side and head-to-tail.
After deriving the quasiparticle excitation spectrum,  we discuss the 
dynamics of the relative phase following a sudden coupling of the independent subsystems. 
{In particular, we calculate the coherence factor of the system, 
which, due to the nonzero tunneling amplitude, it exhibits dephasing-rephasing 
oscillations instead of pure dephasing. These oscillations are} enhanced by a 
higher tunneling energy, and by higher system densities. 
Our predictions provide a benchmark for future experiments at temperatures 
below $T \lesssim 5 \, \mbox{nK}$. 
\end{abstract}

{\it Keywords\/}: Dephasing, rephasing, Bose-Einstein condensation, 
Josephson tunneling, one-dimensional

\maketitle
\section{Introduction} 
The interference of incoherent waves and the emergence of beats are 
fundamental physical phenomena, which can be observed both in classical 
and in quantum systems. 
{Due to their high coherency, and for the 
reduced occupation of excited states, atomic Bose-Einstein condensates (BECs)
are an ideal platform to 
study the dephasing and resynchronization of noninteracting 
modes with commensurate frequencies.} 

{ 
Indeed, since their first experimental realization \cite{cornell1995,davis1995}, 
BECs have represented 
a paradigmatic setup to probe the dynamics of 
macroscopic quantum observables. 
Besides being one of the hallmarks of the transition,
phase coherence is the key property one aims to preserve and it is an ongoing
issue for quantum technologies and devices 
\cite{plenio2017,smerzi2018,trombettoni2018}.
Thus, the question of how long coherence can be sustained is relevant 
for practical applications. 
At the same time, the qualitative and quantitative understanding of 
the concept of phase coherence addresses a fundamental problem 
of many body systems made of interacting constituents \cite{anderson1990}. 

The first theoretical analyses on this topic were devoted to the 
investigation of the tunneling dynamics between independently-formed 
condensates \cite{anderson1990,sols1990,castin1997}. 
By drawing an electrostatic analogy, an effective (semiclassical) theory
can effectively describe the Josephson tunneling between the two atomic systems
\cite{sols1990,sols1998}. A refined description in terms of quasiparticles
can also account for damped oscillations and eventual decoherence effects
due to the presence of an external potential \cite{villain1997,sols2003}. 

Throughout the last two decades, key advancements
have been obtained by focusing on atomic interferometers in lower 
dimensionalities, and adopting refined field-theoretical techniques 
\cite{schumm2005,polkovnikov2006,burkov2007,
gritsev2006,shin2004,shin2005,altman2007,altman2015,
hofferberth2007}. 
Following these works, we consider the paradigmatic experimental 
setup of a pair of one-dimensional parallel quasicondensates 
\cite{schmiedmayer2000,cronin2009}.
This configuration is produced by a fast and coherent splitting of a single 
superfluid with radiofrequency-induced adiabatic potentials 
\cite{zobay2001,schumm2005}. 
Some notable works on this configuration have also shown the emergence 
of a prethermalized stationary state during the relaxation of the closed system \cite{berges2004,gring2012,langen2013,langen2015}. 
}

\begin{figure}
\centering
\includegraphics[scale=0.5]{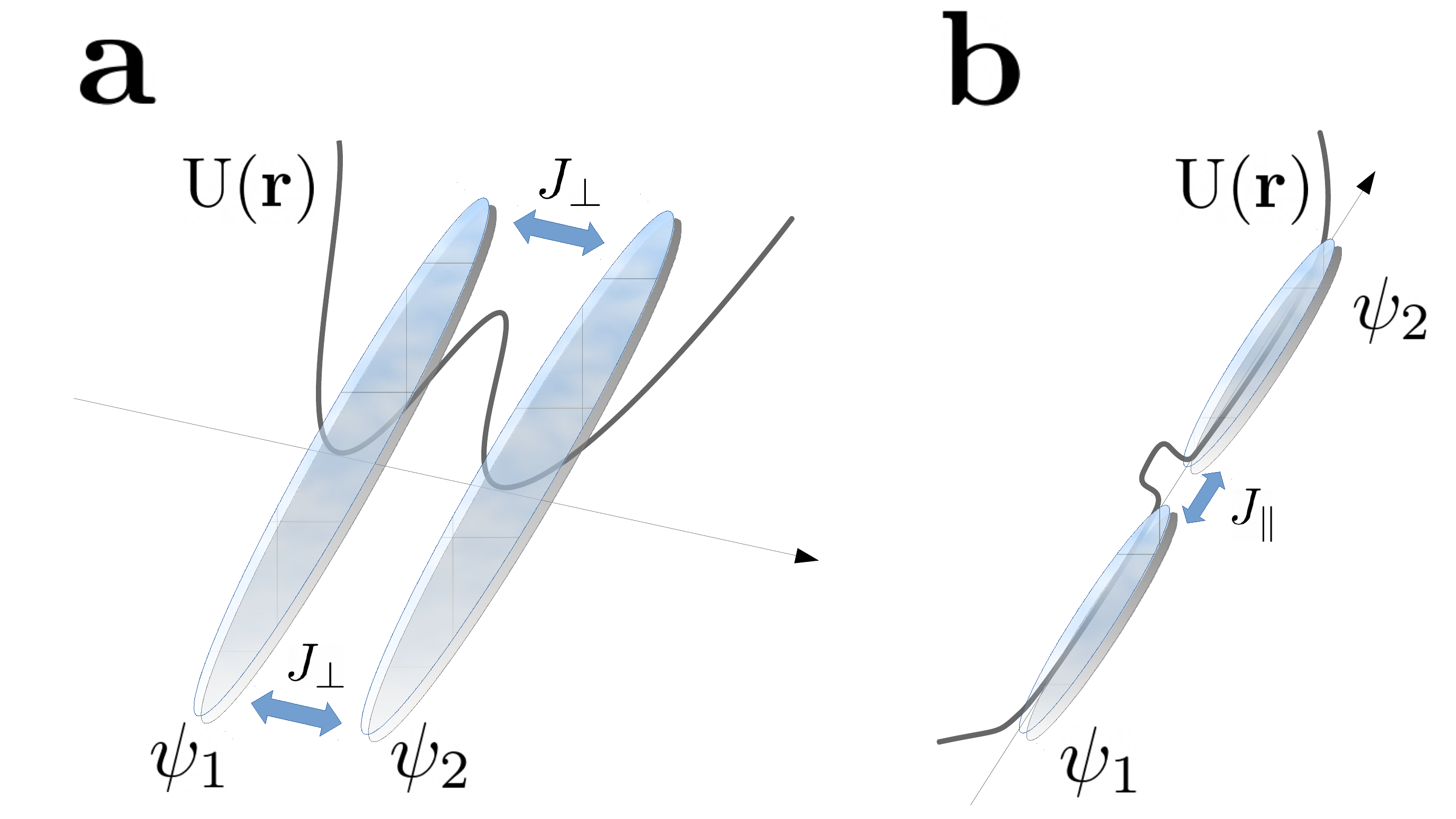}
\caption{Sketch of the system configurations studied in the paper. 
{Depending} on the external potential $U(\vec{r})$, the 
one-dimensional quasicondensates {can be put in a} 
side-by-side, or in a head-to-tail configuration. 
In \textbf{a} the atomic tunneling takes place along the whole 
length of the system, while in \textbf{b} it  only occurs at the center. }
\label{fig1}
\end{figure}

Here we discuss the out-of-equilibrium dynamics of parallel 
quasicondensates in which, after the initial splitting, 
a nonzero tunneling between the subsystems is restored. 
In this context, recent works have analyzed the role of a Josephson 
coupling between the parallel superfluids, considering in particular 
the rephasing dynamics \cite{dallatorre2013,foini2017} 
and the relaxation of the system \cite{foini2015} 
after a quench of the tunneling energy. 
Interestingly, a recent experiment has found a relaxation of the 
Josephson oscillations to a phase-locked equilibrium state 
\cite{pigneur2018}{, an effect which may crucially 
depend on the harmonic confinement \cite{nieuwkerk2019,nieuwkerk2020}.}
In our paper, we investigate both the usual experimental configuration of 
side-by-side parallel quasicondensates with uniform tunneling 
(Fig.~\ref{fig1}a), and the arrangement of head-to-tail parallel 
superfluids with a delta-like Josephson junction (Fig.~\ref{fig1}b) 
\cite{polo2018}. 
In both configurations, due to the similar structure of the 
quasiparticle energy, we find that the coherence factor oscillates in time, 
proving the partial decoherence and resynchronization of the noninteracting 
modes of the coupled quasicondensates. 
This phenomenon is the outcome of the competition between the quantum 
fluctutations of the single quasicondensates, and the coherence-inducing 
Josephson coupling, which introduces a mass gap 
in the quasiparticle spectrum. 
{Previous works on side-by-side superfluids did not predict 
explicitly the oscillations of the coherence factor \cite{foini2015}, 
and our work analyzes the configuration of head-to-tail quasicondensates 
for the first time. 
The experimental observation of these phase oscillations} requires a low initial 
temperature of the quasicondensate ($\sim \mbox{nK}$ range), to avoid 
the intrinsic dephasing induced by thermal fluctuations. 

{
The paper is organized as follows: in Section~\ref{sec2} we introduce 
our model for parallel quasicondensates. 
Then, we reformulate the dynamics of the relative variables in terms 
of quasiparticle excitations in Section~\ref{sec3}, considering both 
the head-to-tail and the side-by-side system configurations. 
Finally, Section~\ref{sec4} explores the phase dynamics of the 
parallel superfluids after a quench in the tunneling amplitude. 
}

\section{The model and the relative dynamics} 
\label{sec2}
In systems of ultracold atoms the tunneling dynamics  is strongly 
dependent on the specific configuration and on the spatial dimension, 
i.e. on the details of the confining potential $U(\vec{r})$. 
Here we discuss the phase dynamics of one-dimensional tunneling 
quasicondensates, which are obtained by confining the 
atoms with radiofrequency-induced adiabatic potentials 
\cite{zobay2001,schumm2005}. 
By properly tuning the trap parameters and the radiofrequency field 
detuning, parallel condensates can be {prepared} 
either in a side-by-side, or in a head-to-tail configuration. 
In both cases, we describe each one-dimensional superfluid with 
{the} complex-valued field $\psi_j(x,t)$, 
where $j=1,2$ labels each subsystem, $x$ is the coordinate 
in the longitudinal direction, and $t$ is  time. 

The real-time Lagrangian $L$ of the system can be written as 
\beqa 
L =  \int_0^L dx \, {\cal L}, \qquad {\cal L} =  
{\cal L}_{tun} + \sum_{j=1,2} \, {\cal L}_{0,j} . 
\label{lagr}
\eeqa
Here ${\cal L}_{0,j}$ is the Lagrangian density of the 
uncoupled quasicondensates, namely 
\beq 
{\cal L}_{0,j} = 
i \hbar \psi_j^* \partial_t \psi_j - {\hbar^2\over 2m} 
|\partial_x \psi_j|^2 - {g\over 2} |\psi_j|^4, 
\eeq
where $m$ is the mass of each atom, $g$ is the one-dimensional interaction 
strength \cite{salasnich}, and $\hbar$ is the reduced Planck constant. 
Note that in our effective one-dimensional approach we omit the external 
potential, assuming that the transverse degrees of freedom are not excited. 
Moreover, in the absence of a longitudinal potential, 
the atomic density along $x$ will be approximately uniform. 
The Lagrangian density ${\cal L}_{tun}$ in Eq.~(\ref{lagr}) describes the 
tunneling of atoms between the two superfluids, namely 
\beq
{\cal L}_{tun} = {J\over 2} \left( \psi_1^*\psi_2 
+ \psi_2^*\psi_1 \right), 
\eeq
where the specific expression of the tunneling energy $J \equiv J(x)$ 
depends on the configuration considered. 
In the side-by-side configuration (Fig.~\ref{fig1}a) we express it as  
$J = J_{\perp}$, 
with $J_{\perp}$ uniform and constant.  In the head-to-tail configuration 
(Fig.~\ref{fig1}b) we choose $J = 2 J_{\parallel} L \, \delta(x)$, 
since tunneling will occur only at the origin 
of the system\footnote{Strictly speaking, in the head-to-tail configuration 
the superfluid {with label} 1 occupies the region 
$ -L \leq x \leq 0$. 
The {invariance under reflection of $x$ in $-x$} 
of ${\cal L}_{0,j}$ allows to write Eq.~(\ref{lagr}).}.

To describe the hydrodynamic properties of the system, we 
{rewrite the complex field as}
$\psi_j(x,t) = [\rho_j(x,t)]^{1/2} \, e^{i \phi_j(x,t)}$ , 
where $\rho_j(x,t)$ is the local density of the bosonic atoms in the 
subsystem $j$ and $\phi_j(x,t)$ is the phase angle \cite{popov1972}.
Substituting this field parametrization into the Lagrangian density 
of Eq.~(\ref{lagr}), we find 
\beqa 
{\cal L} = &&\sum_{j=1,2} \big[ - \hbar \rho_j \, {\dot \phi}_j 
- {\hbar^2\rho_j\over 2m}  \, (\partial_x \phi_j)^2 - 
{\hbar^2\over 8 m \rho_j} \, (\partial_x \rho_j)^2 
- {g\over 2} \rho_j^2  
\big] \nonumber 
\\
&&+ J \sqrt{\rho_1 \rho_2} \, \cos{(\phi_1 - \phi_2)} \; .  
\label{lagr2}
\eeqa 
The tunneling dynamics, as can be seen from the last term of 
Eq.~(\ref{lagr2}), is triggered by a nonzero relative phase 
$\phi_1 - \phi_2$ between the superfluids: our goal is to derive 
an effective Hamiltonian description {of} the relative phase dynamics. 
Indeed, phase correlators are the main observable quantities in 
cold atom interferometry and decoherence experiments 
\cite{cronin2009,langen2013,betz2011}. 
We thus define the total phase $\bar{\phi}$ and relative one $\phi$ as 
\beq
{\bar\phi} = \phi_1 + \phi_2, \qquad \phi = \phi_1 - \phi_2, 
\eeq
and we define the total density $\bar{\rho}$ and the density imbalance 
$\zeta$ as 
\beqa 
{\bar\rho} = \frac{\rho_1 + \rho_2}{2}, \qquad \zeta = { \rho_1 - \rho_2 
\over 2\, {\bar \rho} } \; . 
\eeqa 
After these new variables are substituted in Eq.~(\ref{lagr2}), 
different terms will appear in the new Lagrangian density: 
those which contain only the total (or center of mass) fields, those of the 
relative fields, and the couplings between relative and total fields. 
The latter terms can be neglected under the assumption that the total 
density takes the mean field value of $\bar{\rho}(x,t)=\bar{\rho}$, 
namely that the fluctuations of the total density are negligible, 
and assuming a uniform and constant value of $\bar{\phi}$. 
If the couplings between total and relative modes are neglected, 
we can focus only on the Lagrangian density of the relative modes 
${\cal L}_{rel}$, which reads 
\beqa
{\cal L}_{rel} =&& - \hbar {\bar \rho} \, \zeta {\dot \phi} 
- {\hbar^2{\bar \rho}\over 4m} \, (\partial_x \phi)^2 
- {\hbar^2 {\bar \rho}\over 4 m} \, {(\partial_x \zeta)^2 \over 1 - \zeta^2} 
- g{\bar \rho}^2 \zeta^2 
\nonumber
\\
&& + J \, {\bar \rho} \, \sqrt{1-\zeta^2} \, \cos{(\phi)} \; .
\label{lagr3}
\eeqa 
By considering only the relative fields, we are neglecting in 
particular the anharmonic terms which couple the density 
fluctuations with the phase modes.
These contributions are unimportant in the zero-temperature 
quantum regime that we consider \cite{burkov2007}, but can have 
a relevant role in the dynamics of a system at a  
finite temperature \cite{burkov2007}, 
and for zero interaction between the subsystems \cite{son2002,gallemi2019}. 
In this regard, a discussion of the validity of our scheme for 
the typical experimental parameters will be given in the 
Conclusions. 

Our Lagrangian, Eq.~(\ref{lagr3}), extends the usual two-mode description 
of quantum tunneling by including the contribution of the longitudinal 
excitations of the system. 
This is particularly evident from the Euler-Lagrange equations for 
Eq.~(\ref{lagr3}), which read \beqa 
\hbar {\dot \phi} &=& - J {\zeta \over \sqrt{1-\zeta^2}} \, \cos{(\phi)} 
- 2g {\bar\rho}\, \zeta + {\hbar^2\over 2m} \, 
\bigg[ {\partial_x^2\zeta \over 1 - \zeta^2} 
+ {\zeta \, (\partial_x\zeta)^2 \over (1 - \zeta^2)^2} 
\bigg] \; , 
\label{el1}
\\
\hbar {\dot \zeta} &=& J \sqrt{1 - \zeta^2} \, \sin{(\phi)} 
- {\hbar^2\over 2m} \, \partial_x^2\phi \; .
\label{el2}
\eeqa
These equations reproduce the Josephson-Smerzi equations 
\cite{josephson1962,smerzi1997} for bosons in a double well potential 
if $J = J_0$, with $J_0$ uniform and constant, and the spatial 
dependence of the fields is removed. 
At the same time, our Eqs.~(\ref{el1}), (\ref{el2}) describe the complex 
dynamics in which the tunneling of the {zero momentum} 
mode couples with the Bogoliubov excitations of the superfluids. 

In the linear regime{, for a} small amplitude of the 
relative phase field and of the imbalance, Eqs.~(\ref{el1}), (\ref{el2}) 
can be simplified {as}
\beqa
\hbar {\dot \phi} &=& - (J + 2g {\bar\rho}) \, \zeta 
+ {\hbar^2\over 2m} \, \partial_x^2\zeta  \; ,
\label{el11}
\\
\hbar {\dot \zeta} &=& J \phi - {\hbar^2\over 2m} \, \partial_x^2\phi \;.  
\label{el22}
\eeqa
{These equations can be obtained as} the Euler-Lagrange 
equations of the following low-energy effective Lagrangian 
\beqa 
{\cal L}_{rel} = - \hbar {\bar \rho} \, \zeta {\dot \phi} 
- {\hbar^2{\bar \rho}\over 4m} \, (\partial_x \phi)^2 
- {\hbar^2 {\bar \rho}\over 4 m} \, (\partial_x \zeta)^2 
- {(2g{\bar \rho}^2+J{\bar\rho})\over 2} \, \zeta^2 
- {J {\bar \rho} \over 2} \, \phi^2, 
\label{lagr4}
\eeqa
which is quadratic in the relative fields.

\section{{Reformulation in terms of quasiparticles}}
\label{sec3}
In this Section, we derive the excitation spectrum of the coupled superfluids 
in the side-by-side configuration, and in the head-to-tail one. 
For both configurations, we reformulate the complex dynamics of {interacting} tunneling 
quasicondensates in terms of noninteracting 
quasiparticle excitations. 

\subsection{Side-by-side parallel quasicondensates}
{To describe the side-by-side tunneling configuration 
(Fig.~\ref{fig1}a), 
we consider a uniform and constant tunneling energy $J=J_{\perp}$.}
Let us introduce the Fourier representation of 
the relative phase $\phi(x,t)$ and of the imbalance $\zeta(x,t)$, namely 
\beqa
\phi(x,t) = {\sqrt{2}\over L}\sum_{k \geq 0} \phi_k(t) \, \cos{(k x)},  
\quad 
\phi_k(t) = \alpha_k \int_0^L dx \, \phi(x,t) \cos{(k x)},
\\
\zeta(x,t) = {\sqrt{2}\over L}\sum_{k \geq 0} \zeta_k(t) \, \cos{(k x)}, 
\quad 
\zeta_k(t) = \alpha_k \int_0^L dx \, \zeta(x,t) \cos{(k x)},
\eeqa
where we have imposed open boundary conditions, i.e. $\partial_x\phi(0,t)=
\partial_x\phi(L,t)=0$, $\partial_x\zeta(0,t)=
\partial_x\zeta(L,t)=0${,} at any time $t$. 
The wavevector $k$ is given by $k=\pi n/L$, with $n=0,1,2,...$ and 
we define the parameter $\alpha_k = \sqrt{2}$ for $k \neq 0$, 
while $\alpha_0 = 1/\sqrt{2}$.

We now substitute these Fourier field decompositions into the 
real space Lagrangian $L_{rel}$ associated to the Lagrangian density 
${\cal L}_{rel}$ of Eq.~(\ref{lagr4}). 
Thereafter, the Euler-Lagrange equation for $\phi_k$ yields an explicit 
expression of the imbalance mode $\zeta_k$: substituting it in the 
Lagrangian $L_{rel}$, we get an effective Lagrangian for 
the phase modes only. 
With the usual Legendre transform, we can write the corresponding 
effective Hamiltonian as
\beq 
H =  \sum_k \left[ 
{p_k^2\over 2M_k} + {M_k \over 2} \omega_k^2 \phi_k^2 \right] \; , 
\label{ham1}
\eeq
where $p_k=M_k {\dot\phi}_k$ are the linear momenta 
associated to the generalized coordinates $\phi_k$. 
In this way, we have reformulated the low-energy description of tunneling 
quasicondensates as a sum of noninteracting harmonic oscillators, 
or quasiparticles. Each oscillator has a wavevector-dependent 
effective mass $M_k$, namely  
\beq
M_k = \frac{\hbar^2 {\bar \rho}}{L}  {1\over { (J_{\perp} + 2g {\bar\rho}) 
+ \hbar^2k^2/(2m) }}, 
\label{masssbs}
\eeq
and a quasiparticle energy given by 
\beq 
\hbar \omega_k  = \sqrt{ J_{\perp}(J_{\perp} + 2g {\bar\rho}) + 
{\hbar^2k^2\over 2m}\left( {\hbar^2k^2\over 2m} + 
2(J_{\perp} + g {\bar\rho}) \right) } \; .
\label{elementaryexc}
\eeq
This quasiparticle energy is a gapped Bogoliubov-like spectrum, 
{and the zero-mode term reproduces the usual Josephson 
tunneling energy.} 
Indeed, for $k=0$, the excitation energy of Eq. (\ref{elementaryexc}) reads 
\beq 
\hbar \omega_0 = \sqrt{ J_{\perp}^2 + {2 g \bar{\rho} J_{\perp}}}	\, , 
\label{josephson}
\eeq
{and $\omega_0$ is exactly} the Josephson oscillation 
frequency in the absence of spatial dependence {of the fields} 
\cite{smerzi1997}. 
At the same time, in the absence of tunneling among the subsystems, i.e.
setting $J_{\perp}=0$, we get 
\beq 
\hbar \omega_{k,B}  = \sqrt{{\hbar^2k^2\over 2m}
\left( {\hbar^2k^2\over 2m} + 2g {\bar\rho} \right) } \; ,   
\label{bogoliubov}
\eeq
that is the familiar Bogoliubov spectrum 
of elementary excitations along the longitudinal direction 
\cite{bogoliubov1947}. 
For small wavevectors, the Bogoliubov spectrum {can be approximated by} 
the phonon spectrum $\hbar\omega_k =  c_s \, \hbar k$, with 
the speed of sound given by $c_s=(g{\bar\rho}/m)^{1/2}$. 
We emphasize that, in this low-energy non-tunneling regime 
the quasiparticle mass is wavevector independent, namely $M=\hbar^2/(2gL)$, 
and our procedure reproduces the results of Ref.~\cite{altman2007}. 
Moreover, for a nonzero tunneling energy $J_{\perp}$ between the 
quasicondensates{, and 
in the Josephson regime of $g \bar{\rho} \gg J_{\perp}$}, our excitation spectrum of Eq.~(\ref{elementaryexc}) 
is consistent with the one derived in Refs.~\cite{foini2017,foini2015}. 
{With respect to these works, therefore, 
we obtain a more general excitation spectrum, which includes the 
free-particle behavior of the Bogoliubov spectrum for large wavevectors.}

\subsection{Head-to-tail parallel quasicondensates}
In head-to-tail parallel superfluids {(Fig.~\ref{fig1}b)} 
we model the tunneling energy as 
$J(x)=2 J_{\parallel} L \delta(x)$. 
{We implement a low-energy effective description of 
this system configuration} 
by neglecting the spatial derivatives of the imbalance in Eq.~(\ref{el11}). 
For consistency with this approximation, we work in the Josephson regime 
of $g {\bar\rho} \gg J_{\parallel}$, in which the experiments are usually 
performed \cite{pigneur2018,betz2011}. 
In this case, Eq.~(\ref{el11}) can be easily solved as 
$\zeta = - \hbar \dot{\phi}/(2 g \bar{\rho})$, and, substituting it 
into Eq.~(\ref{lagr4}), we get an effective Lagrangian density 
{for the relative phase} 
\beqa 
\mathcal{L}_{rel} =  
{\hbar^2\over 4g} \, {\dot \phi}^2  
- {\hbar^2{\bar \rho}\over 4m} \, (\partial_x \phi)^2 
- {J_{\parallel}{\bar \rho}} 
\,  L \delta(x) \, \phi^2 . 
\label{lagranna}
\eeqa

In analogy with the previous subsection, {here we} 
decompose the phase field as 
\beq 
\phi(x,t) = {1\over \sqrt{L}} \sum_n q_n(t) \ \phi_n(x) \; , 
\label{phidecomposition}
\eeq
where the $\phi_n(x)$ are real and {orthonormal} eigenfunctions 
of the eigenvalue problem 
\beq  
\left[ - {\hbar^2\over 2m} \partial_x^2 + 2 J_{\parallel} L \delta(x) 
\right] \phi_n(x) = \epsilon_n \, \phi_n(x) \; 
\label{eigenvalueproblem}
\eeq
with open 
boundary conditions $\partial_x\phi_n(0)=\partial_x\phi_n(L)=0$. 
The eigenvalues $\epsilon_n$ of Eq.~(\ref{eigenvalueproblem}) are 
determined by the following 
equation for $\epsilon$ 
\beq 
\sqrt{2m L^2 \epsilon \over \hbar^2 } \; 
\tan{\left( \sqrt{2m L^2 \epsilon \over \hbar^2 }\right)} 
= {2m L^2 J_{\parallel} \over \hbar^2} ,
\label{epsilon}
\eeq
which admits an infinite set of solutions, labelled by the integer 
number $n$ \cite{atkinson1975}. 

{ These solutions may either be obtained numerically 
or well approximated by the following analytical approximations, 
holding when $n \ll \tilde{J_{\parallel}} $ or 
$n > \tilde{J_{\parallel}} $, where 
$\tilde{J_{\parallel}} = J_{\parallel} \pi / ( \hbar^2 \pi^2 / (2 m L^2) )$.
First of all, in the regime of 
$n > \tilde{J_{\parallel}} $,}  
the solutions $\epsilon_n$ of Eq.~(\ref{epsilon}) 
{are well approximated by }
\beq 
\epsilon_n = \frac{1}{4}  \bigg(     \sqrt{ 4 J_{\parallel} 
+  \frac{\hbar^2\pi^2n^2}{2mL^2} }  
+ \sqrt{\frac{\hbar^2\pi^2 n^2}{2mL^2}  } \bigg)^2 , 
\label{perturbativespectrum0}
\eeq 
which{, essentially,} is the energy of a 
{free particle} with a mass gap. 
{ Actually, provided that $\tilde{J_{\parallel}} \le \pi/16$, 
Eq.~(\ref{perturbativespectrum0}) holds also for $n=0$.
In the opposite case of $n \ll \tilde{J_{\parallel}}$ the solutions 
of Eq.~(\ref{epsilon}) are instead well approximated by:
\beq
\epsilon_n = \frac{\hbar^2 \pi^2}{2 m L^2}
\frac{J_{\parallel}}{J_{\parallel}+ \frac{\hbar^2 }{2 m L^2}} 
\bigg(n+\frac{1}{2}\bigg)^2.
\label{delta2}
\eeq
which holds also for $n=0$ when $\tilde{J_{\parallel}} \ge 1 / \pi$. 
Finally, we emphasize that in the intermediate range of values of 
$ \tilde{J_{\parallel}}$, namely for $ \pi /16 \leq \tilde{J_{\parallel}} \leq 1/\pi$, 
the lowest solution ($n=0$) of Eq. (\ref{perturbativespectrum0}) may be written as 
$ \epsilon_0 =  \hbar^2 \pi^2/(32 m L^2)$. 
}

In analogy with the the previous subsection, we insert the decomposition 
of Eq.~(\ref{phidecomposition}) into the effective Lagrangian density 
$\mathcal{L}_{rel}$ of Eq.~(\ref{lagranna}){. Then,} 
calculating the corresponding Hamiltonian with a Legendre transformation, 
we get 
\beq 
H = \sum_n \left[ {P_n^2\over 2M} + 
{M\over 2} \Omega_n^2 \phi_n^2 \right] \; , 
\label{ham2}
\eeq
where $P_n=M {\dot \phi}_n$ are generalized momenta associated to the 
phase modes $\phi_n$. Again, we are describing the dynamics of the 
relative degrees of freedom for head-to-tail parallel quasicondensates 
as noninteracting harmonic oscillators with the effective mass 
$M=\hbar^2/(2gL)$, and the harmonic frequency 
$\Omega_n = (2 g \bar{\rho} \, \epsilon_n  / \hbar^2)^{1/2}$. 
From the knowledge of the eigenvalues $\epsilon_n$, obtained from 
Eq.~(\ref{epsilon}), one immediately derives the frequencies $\Omega_n$. 
In the following {Section} we will discuss the relative 
phase dynamics of this system 
in a regime of {very} small Josephson tunneling 
$J_{\parallel}${, where} 
Eq.~(\ref{perturbativespectrum0}) is reliable  {even for $n=0$}. 
In this case, the quasiparticle energies can be expressed as 
\beq 
\hbar\Omega_k =  \sqrt{2 g \bar{\rho} \, J_{\parallel} 
+ \frac{g \bar{\rho}}{2} \, \frac{\hbar^2k^2}{2m} }  
+ \sqrt{\frac{g \bar{\rho}}{2} \, \frac{\hbar^2k^2}{2m}  }   , 
\label{perturbativespectrum}
\eeq
where, {as in} the previous subsection, we have introduced 
the wavevector $k=n \pi/L$. 

\section{Phase oscillations in one-dimensional tunneling quasicondensates}  
\label{sec4}
We now discuss the phase dynamics after a quantum quench of the 
tunneling amplitude, {describing} a procedure which 
can be implemented 
both for tunneling side-by-side and head-to-tail quasicondensates. 

The {time evolution} of the relative phase and of its 
correlation functions is crucially dependent on the experimental 
protocol adopted to prepare the initial 
state{, and on the Hamiltonian under which the system evolves.} 
Here we suppose that the atomic system, initially confined in a 
one-dimensional single quasicondensate, is symmetrically split 
into a couple of parallel superfluids. In the experiments, 
the splitting procedure consists in tuning a radiofrequency field 
to create a double well adiabatic potential for the parallel condensates 
\cite{schumm2005}. 
For a large enough number of atoms in the system, as argued in Refs. \cite{burkov2007,altman2007,foini2015}, the splitting procedure leads 
to a Gaussian probability distribution of the imbalance, with zero mean 
and a variance proportional to the mean-field density $\bar{\rho}$. 
Hence, assuming a minimum uncertainty for the relative phase $\phi$, 
canonically conjugated to the imbalance $\zeta$, the initial wavefunction 
in the relative phase representation is given by \cite{altman2007} 
\beq 
\Psi[\{\phi_k\},t=0]  
\simeq 
\prod_k \Psi_k(\phi_k,t=0), \, \, \, \quad  
\eeq
where the single phase wavefunction reads 
\beq
\Psi_k(\phi_k,t=0) = {1\over \pi^{1/4} \sigma^{1/2}} \, 
e^{-\phi_k^2\over 2\sigma^2} \; ,
\label{Psitot}
\eeq
with $\sigma^2=L^2/N=L/\bar{\rho}$, since $\phi_k$ 
is dimensionally a length. 
Note that, in order to have small fluctuations in the relative phase field, 
the splitting time $\tau_s$ must be very short. In particular, $\tau_s$ 
must satisfy the condition of $\tau_s \ll \xi / c_s$, with 
$\xi=(\hbar^2/mg\bar{\rho})^{1/2}$ the healing length, 
and $c_s$ the speed of sound of the single superfluid \cite{altman2007}. 
At the same time, $\tau_s$ must be long enough to avoid the excitation of 
the transverse modes of the one-dimensional superfluids. 

Given the Gaussian initial state in terms of the phase modes $\phi_k$, 
we want to calculate its time evolution under the harmonic Hamiltonians 
of the tunneling quasicondensates, namely Eq.~(\ref{ham1}), (\ref{ham2}). 
{Experimentally, the time evolution under these Hamiltonians 
can be implemented by lowering the barrier which separates the superfluids. 
This} procedure must take place in a finite but quick enough time 
$\tau_J \approx \tau_s$, to avoid excessive dephasing on one hand 
\cite{altman2007}, and to keep the system near the initial state, 
Eq.~(\ref{Psitot}), on the other.  

In the Schr\"odinger picture, the quantum dynamics of the system of 
oscillators with Hamiltonians of Eqs.~(\ref{ham1}), (\ref{ham2}), 
follows the Schr\"odinger equation 
\beq 
i \hbar \partial_t \Psi_k(\phi_k,t) = 
\left[ -{\hbar^2\over 2M_k} 
{\partial^2\over \partial \phi_k^2} + {M_k\omega_k^2\over 2} \phi_k^2 
\right] \Psi_k(\phi_k,t), 
\label{tdse}
\eeq
where the mass $M_k$ and the oscillator frequency $\omega_k$ are those 
calculated previously in the side-by-side, 
and in the head-to-tail configurations. 
For each oscillator, the wavefunction remains Gaussian during the 
time evolution, but the standard deviation evolves with time. 
In particular, the probability density of the mode $k$ at time $t$ becomes 
\cite{robinett2004}
\beq 
|\Psi_k(\phi_k,t)|^2 = {1\over \pi^{1/2} \sigma_k(t)}
e^{-{\phi_k^2\over \sigma_k^2(t)}} \; 
\eeq
with {$\sigma_k^2(t)$} given by 
\beq 
\sigma_k^2(t) = \sigma^2 \cos^2(\omega_k t)+ 
\left({\hbar^2\over M_k^2 \omega_k^2 \sigma^2}\right) 
\sin^2(\omega_k t) \; . 
\label{sigma}
\eeq 
Thus, during the time evolution, the standard deviations of the various 
modes evolve differently one from the other, since their frequencies 
$\omega_k$ and effective masses $M_k$ are different.

The phase coherence of the system is quantified by the coherence factor 
\cite{altman2007,hofferberth2007}
\beq
{\cal C}(t)=\langle \cos(\phi) \rangle_t = 
e^{-{1\over 2 L^2}\sum_k \langle \phi_k^2 \rangle_t} = 
e^{-{1\over 4 L^2}\sum_k  \sigma_k^2(t)} 
\label{coherence}
\eeq
where the averaging is calculated over the wavefunction at time $t$, 
namely $\Psi[\{\phi_k\},t]$ in the $\phi_k$ representation, 
and the second equality holds for Gaussian distributed variables 
\cite{altman2007,kardar}. 
As Eq.~(\ref{coherence}) shows, the coherence factor evolves in time 
as a function of the sum over all $\sigma_k^2(t)$. 
Depending on the specific form of the excitation spectrum $\omega_k$ 
and on the oscillator mass $M_k$, ${\cal C}(t)$ will show a different 
qualitative behavior. 
For non-tunneling parallel quasicondensates with a phononic spectrum, 
and a $k$-independent mass, the coherence factor decays exponentially to zero 
due to the dephasing of the modes \cite{altman2007}. 
The dephasing of ${\cal C}(t)$ is however absent in tunneling superfluids, 
where the nonzero tunneling energy $J$ introduces a mass gap 
in the excitation spectrum \cite{foini2017}. 
This is exactly what happens in our case, as can be seen in the 
quasiparticle spectra of Eqs.~(\ref{elementaryexc}), 
and (\ref{perturbativespectrum}). 

In the next subsections we will explicitly discuss our results for ${\cal C}(t)$, 
obtained in the side-by-side, and in the head-to-tail configurations. 
Before of that, let us briefly remind how the coherence factor is measured. 
In the experiments, after releasing the external potential which confines 
the quasicondensates, one can measure the distribution of the relative phase 
$\phi$ from the interference pattern between the subsystems. 
Knowing the relative phase profile allows to calculate the coherence factor 
of Eq.~(\ref{coherence}), by integrating $\mbox{e}^{i \phi}$ over the length 
of the imaging system \cite{hofferberth2007,nieuwkerk2018}.

\begin{figure}
\centering
\includegraphics[scale=1.0]{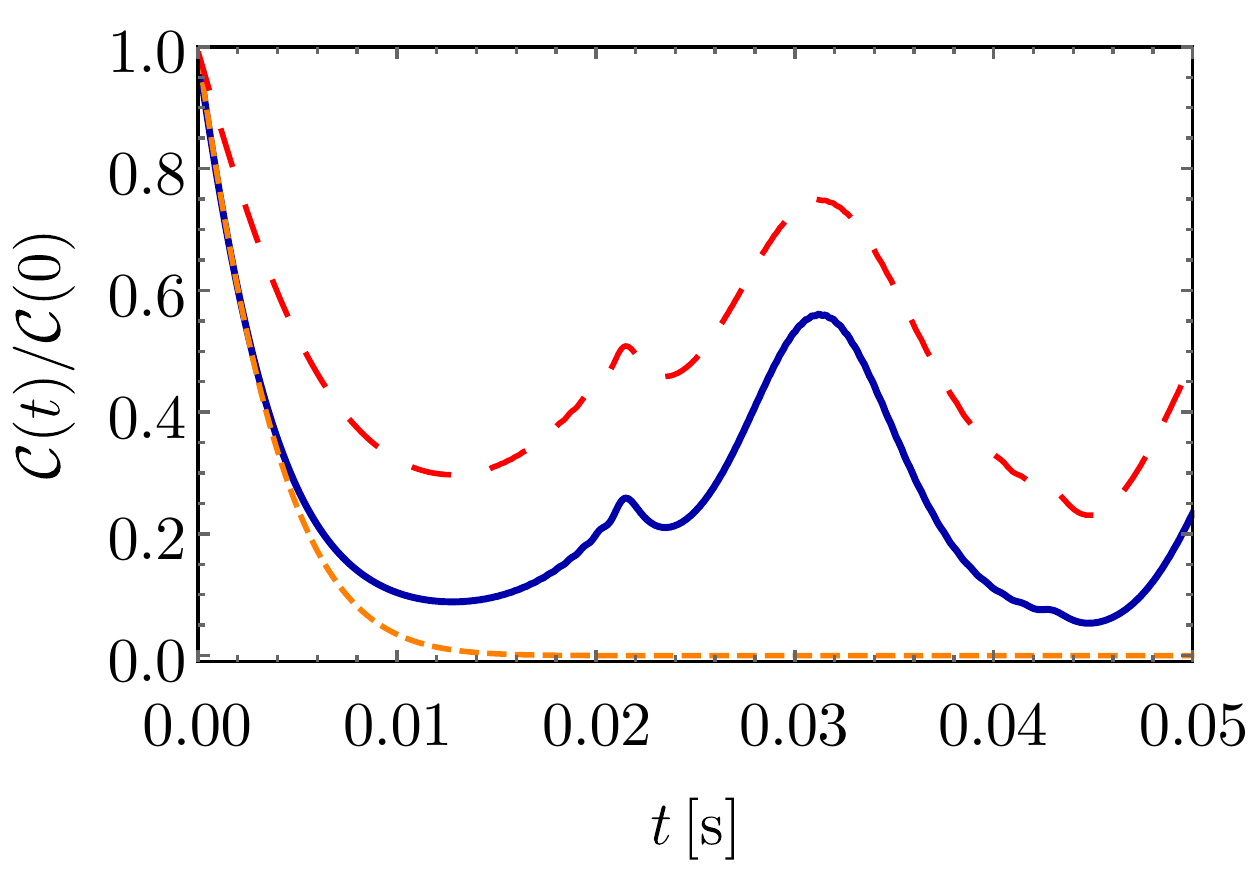}
\caption{Recurrent dephasing-rephasing dynamics of side-by-side 
tunneling quasicondensates for different system lengths: 
$L=50 \, \mu \mbox{m}$ (blue solid line), $L=100 \, \mu \mbox{m}$ 
(red long-dashed line). 
Differently from uncoupled superfluids, which dephase exponentially 
\cite{altman2007} (orange dashed line), 
{in the tunneling configuration a clear oscillation of 
the phase coherence appears,
as a result of} the partial resynchronization of the quasiparticle oscillators 
(see also Fig.~\ref{fig3}). 
{
The parameters adopted here are $J_{\perp}/\hbar= 5/(2 \pi) \, \mbox{Hz}$, 
${\bar \rho}=50 \, \mu \mbox{m}^{-1}$, $^{87}$Rb mass, 
$g=g_{3D}/(2\pi l^2)$, with $l=\sqrt{\hbar/(m \Omega)}$, 
and $\Omega=2 \pi \times 2.1 \, \mbox{kHz}$.} }
\label{fig2}
\end{figure}
\begin{figure}
\centering
\includegraphics[scale=1.0]{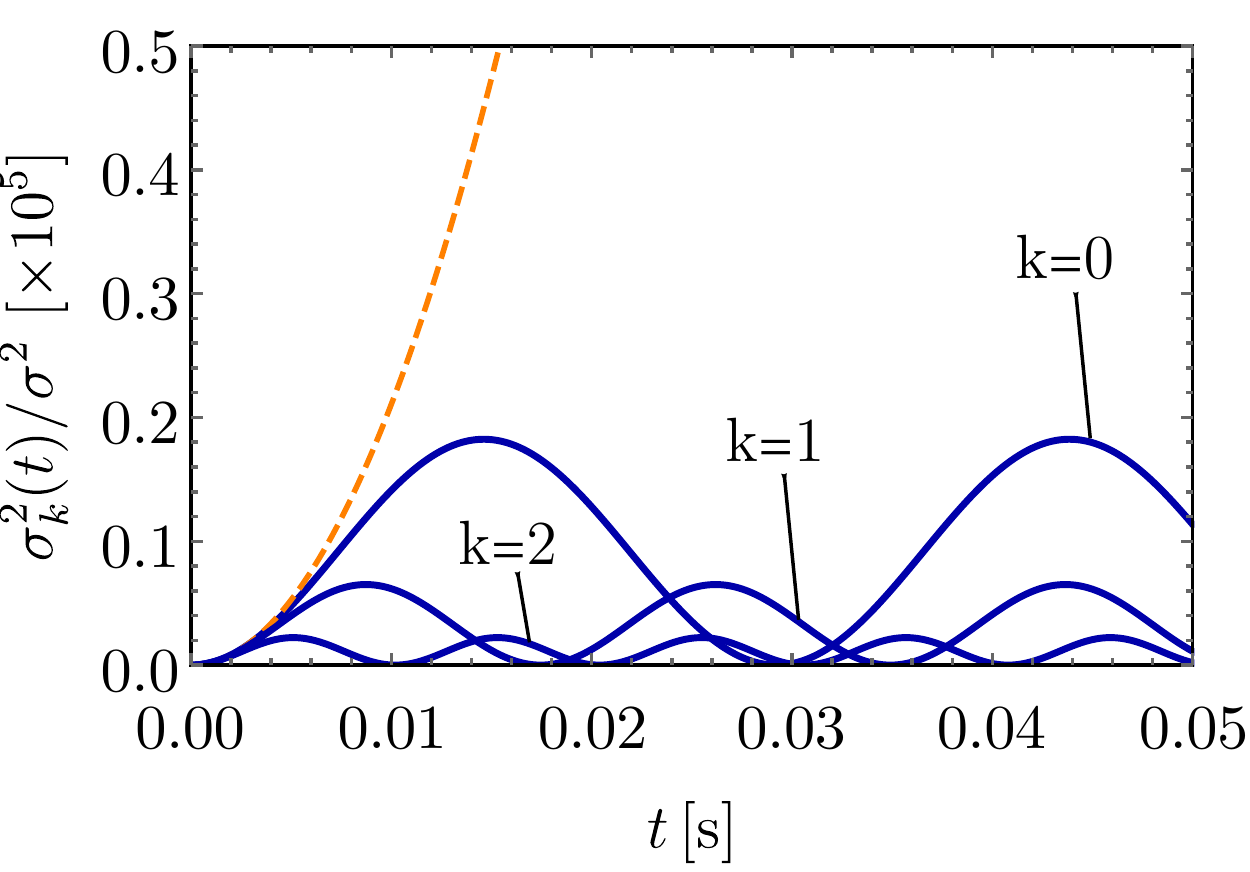}
\caption{Time evolution of the Gaussian standard deviations $\sigma_k^2(t)$, 
rescaled with the $\sigma^2$ of the initial state{, 
in the side-by-side configuration}. 
In the absence of Josephson tunneling, the $\sigma_0^2(t)$ increases 
quadratically with time (orange dashed line), 
producing dephasing \cite{burkov2007}. 
{If tunneling occurs the mode $\sigma_0^2(t)$ does 
not diverge in time, and dephasing does not occur. Note that the modes 
here represented (blue solid lines) have different frequencies, and decreasing 
amplitudes for increasing $k=\{0,1,2\} \, \pi/L$.} 
In this plot, we use the same parameters of the blue solid line in 
the previous figure.}
\label{fig3}
\end{figure}
\subsection{Side-by-side parallel quasicondensates}
The phase coherence ${\cal C}(t)$ of side-by-side quasicondensates 
{is shown in Fig.~\ref{fig2}. 
It is obtained} from Eqs.~(\ref{sigma}), 
(\ref{coherence}) using the gapped Bogoliubov-like spectrum 
of Eq.~(\ref{elementaryexc}), and the quasiparticle mass 
of Eq.~(\ref{masssbs}). 
Differently from non-tunneling superfluids (orange dashed line), 
we find a clear rephasing phenomenon, which is more evident 
{in} longer systems (red line, $L=100 \, \mu \mbox{m}$) 
than in shorter ones (blue solid line, $L=50 \, \mu \mbox{m}$). 
We find that the oscillations of ${\cal C}(t)$ become more frequent 
and with higher amplitude if the atomic density $\bar{\rho}$ is 
increased. 
A similar behavior is obtained also if the value of $J_{\perp}$ is 
increased, signalling that the phase coherence is enhanced by a 
stronger tunneling between the subsystems. 

The substructures around $2 \, \mbox{ms}$, and $4 \, \mbox{ms}$ 
in the curves of Fig.~\ref{fig2} are due to the incoherent sum of the 
Gaussian standard deviations $\sigma_k^2(t)$, which appear at the exponential 
of the coherence factor ${\cal C}(t)$. 
This can be seen in Fig.~\ref{fig3}, where we plot the 
normalized standard deviations $\sigma_k^2(t)/\sigma^2$ of the Gaussian 
wavefunctions $\Psi_k(\phi_k,t)$, for $k=\{0,1,2\} \, \pi/L$ 
(we use $k$ instead of $n$ with a slight abuse of notation). 
Due to the different quasiparticle energy $\hbar \omega_k$ for each mode, 
the deviations $\sigma_k^2(t)/\sigma^2$ oscillate with different frequencies, 
and with smaller amplitudes for larger values of $k$. 
The sum over all these modes produces the dephasing-rephasing oscillations 
shown in Fig.~\ref{fig2}. 
This behavior of the system is completely different from that of uncoupled 
condensates, in which, since $\sigma_0^2(t) \propto t^2$ 
(orange dashed line of Fig.~\ref{fig3}), the coherence factor decays 
exponentially to zero \cite{altman2007}. 

We point out that, following Refs.~\cite{altman2007,foini2017,foini2015}, 
in {the evaluation of} the sum over the wavevectors $k$ of Eq.~(\ref{coherence}) 
we have introduced a natural cutoff 
$\Lambda=\pi/\xi$, where $\xi=\hbar/\sqrt{mg{\bar \rho}}$ 
is the healing length. 
Considering that the {$\sigma_k(t)$ strongly decrease 
in amplitude if $k$ is increased}, 
the cutoff does not introduce any {spurious} unphysical behavior.

We emphasize that, while our zero-temperature theory predicts 
the phase oscillations to continue indefinitely, 
a high phase coherence cannot be observed in the experiments 
for very long times. 
This is due to the intrinsic presence of thermal excitations, 
and the conditions under which our theory is valid 
are highlighted in the Conclusions.

\begin{figure}[hbtp]
\centering
\includegraphics[scale=1.0]{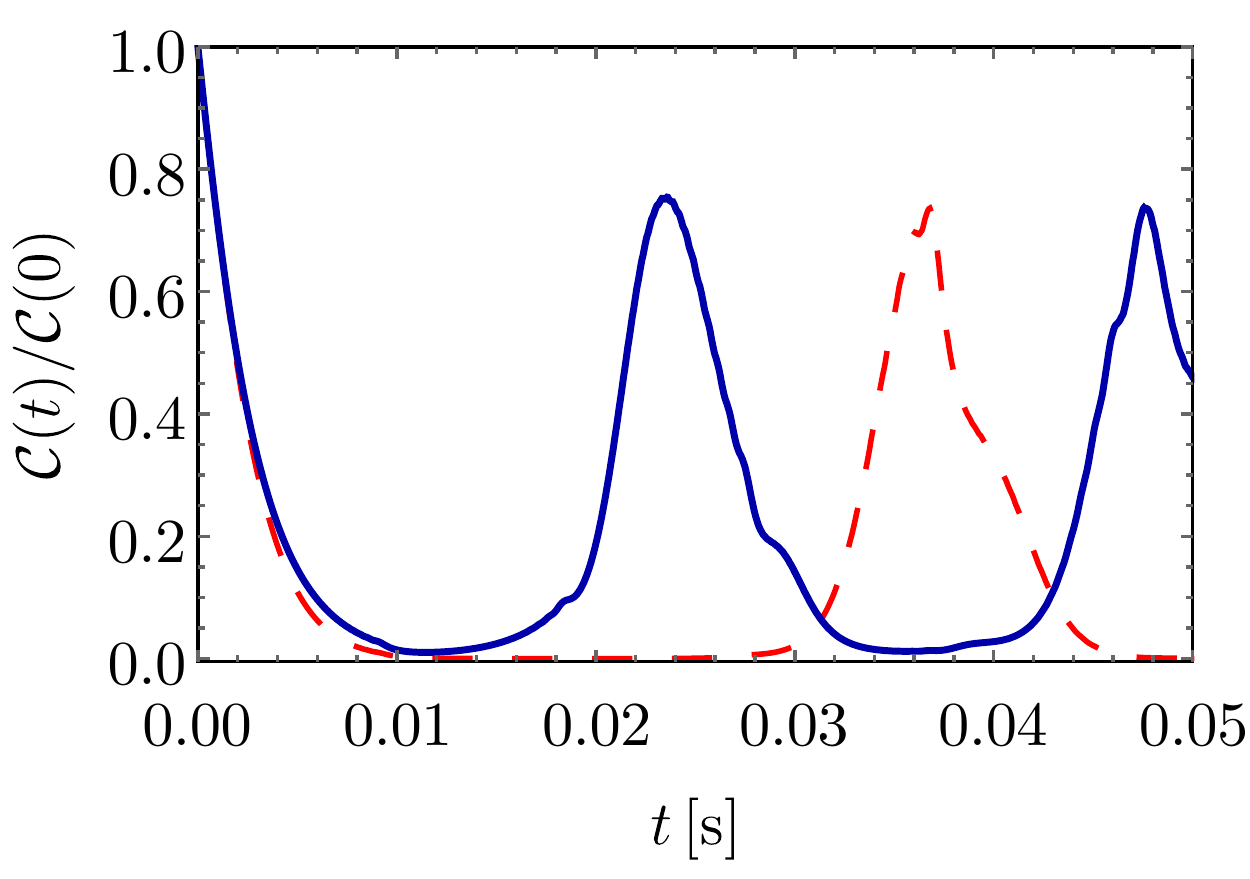}
\caption{{Oscillations of the coherence factor ${\cal C}(t)$ 
in head-to-tail parallel quasicondensates, where 
$J_{\parallel}/\hbar = 8/(2\pi) \, \mbox{Hz}$ 
(blue solid line), and $J_{\parallel}/\hbar = 2/(2\pi) \, \mbox{Hz}$ (red dashed line). 
Note that a higher value of the tunneling energy} $J_{\parallel}$ 
enhances the frequency of the oscillations, and increases the overall coherence. {The coherence factor is calculated by using Eq.~(\ref{coherence}), with the excitation spectrum $\omega_k$ given by the numerical solution of Eq.~(\ref{epsilon}).}
For both curves, we use 
$L=30 \, \mu\mbox{m}$, ${\bar \rho}=100 \, \mu \mbox{m}^{-1}$, 
and the other parameters as in Fig.~\ref{fig2}. }
\label{fig4}
\end{figure}
\subsection{Head-to-tail parallel quasicondensates}
The low-energy excitation spectrum of head-to-tail parallel superfluids, 
Eq.~(\ref{perturbativespectrum}), is qualitatively similar to that of 
side-by-side ones, Eq. (\ref{elementaryexc}). 
Due to this formal analogy, we expect to observe a similar qualitative 
picture in the relative phase dynamics of the system. 
Indeed, after a sudden coupling of the quasicondensates with a delta-like 
Josephson junction with tunneling energy $J_{\parallel}$, 
the coherence factor ${\cal C}(t)$ of Eq.~(\ref{coherence}) oscillates 
in time. 
{As in the previous Section, the oscillation is a result of} 
the partial dephasing and rephasing of the quasiparticle modes. 
In Fig.~\ref{fig4} we plot ${\cal C}(t)$ as a function of time for different  
tunneling energies:  
$J_{\parallel}/\hbar = 8/(2\pi) \, \mbox{Hz}$ (blue solid line) and 
$J_{\parallel}/\hbar = 2/(2\pi) \, \mbox{Hz}$ (red dashed line). 
Thus, for higher values of $J_{\parallel}$ the oscillations 
of ${\cal C}(t)$ are more frequent, and with a higher baseline, 
signalling a higher phase coherence. 
As we stress in the Conclusions, and similarly to the case of head-to-tail 
quasicondensates, our zero-temperature approach does not describe the thermal 
decoherence processes which could dissipate these oscillations at long times. 
Thus, in the experiments, a sufficiently high value of $J_{\parallel}$ 
is needed, to observe the phase oscillations before of the onset of 
thermal dephasing. 

Let us finally note that a single quasicondensate can be split into a 
couple of parallel head-to-tail superfluids by superimposing an 
optical potential to the longitudinal magnetic trap \cite{tajik2019}. 
The tunability of this configuration and the possibility to engineer 
a box-like potential along $x$ are useful tools to satisfy the 
hypothesis of a constant density $\bar{\rho}$, and to implement 
our open boundary conditions.

\section{Conclusions}
We have derived the excitation spectrum of tunneling quasicondensates 
in two distinct experimental configurations: side-by-side, 
and head-to-tail superfluids. 
We find that the sudden coupling of the independent quasicondensates 
produces an interesting nonequilibrium dynamics 
of the relative phase between the subsystems. 
In both system configurations, due to the formal similarities 
in the excitation spectrum, the coherence factor oscillates in time, 
driven by a nonzero tunneling between the quasicondensates. 
The coherence and the frequency of the oscillations are enhanced 
for higher atomic densities and for higher tunneling energies. 

The unavoidable thermal fluctuations of temperature $T$ setup a 
{typical time $\tau$} above which our zero-temperature 
results are no more reliable and a thermal-induced dephasing is expected. 
The microscopic source of the decoherence is the coupling 
between the center of mass modes, constituting a thermal bath, and 
the collective modes of the relative phase \cite{burkov2007}. 
We estimate the {typical time $\tau$ as 
$\tau = \hbar/(k_B T)$,} where $k_B$ is the Boltzmann constant 
\cite{burkov2007}.
Thus, our analytical and numerical dynamical results become unreliable
for a time duration $t$ such that {$t \gg \tau$.} 
Considering this criterion, the experimental observation of the phase 
oscillations shown in Figs.~\ref{fig2}, and \ref{fig4}, requires that the 
temperature before the splitting is in the $ \mbox{nK}$ range. 
{Indeed, assuming that the oscillations are dissipated 
only in part on a time interval up to $10 \, \tau$, 
an initial temperature of $5 \, \mbox{nK}$ would allow to observe 
$15 \, \mbox{ms}$ of the dynamics. 
By properly tuning the tunneling energy, this time should be sufficient 
to observe at least a partial rephasing of the system.} 

We are fully aware that reaching these conditions of temperature in our 
setup poses a real experimental challenge, since analogous experiments 
in atom chips are currently performed at a temperature of 
$18 \, \mbox{nK}$ \cite{pigneur2018}. 
At the same time, considering the huge experimental developments of 
the last decades, we believe that the implementation of 
these experimental conditions will be soon technically feasible. 

\ack
The authors thank Luca Dell'Anna,  
Anna Minguzzi, and Alessio Notari for useful discussions, Ehud Altman, 
Juan Polo, Marine Pigneur and J\"org Schmiedmayer for 
enlightening e-mails. 
LS acknowledges the BIRD Project 
``Time-dependent density functional theory of quantum atomic mixtures'' 
of the University of Padova for partial support.

\end{document}